# Bi$_2$O$_2$Se nanowires presenting high mobility and strong spin-orbit coupling

Kui Zhao[1,2], Huaiyuan Liu[1,2], Congwei Tan[3], Jianfei Xiao[1,2], Jie Shen[1,4], Guangtong Liu[1,4], Hailin Peng[3,a)], Li Lu[1,2,4,a)], and Fanming Qu[1,2,4,a)]

[1]*Beijing National Laboratory for Condensed Matter Physics, Institute of Physics, Chinese Academy of Sciences, Beijing 100190, China*

[2]*School of Physical Sciences, University of Chinese Academy of Sciences, Beijing 100049, China*

[3]*Center for Nanochemistry, Beijing Science and Engineering Center for Nanocarbons, Beijing National Laboratory for Molecular Sciences, College of Chemistry and Molecular Engineering, Peking University, Beijing 100871, China*

[4]*Songshan Lake Materials Laboratory, Dongguan, Guangdong 523808, China*

[a)] Authors to whom correspondence should be addressed: hlpeng@pku.edu.cn; lilu@iphy.ac.cn; fanmingqu@iphy.ac.cn

**Systematic electrical transport characterizations were performed on high-quality Bi$_2$O$_2$Se nanowires to illustrate its great transport properties and further application potentials in spintronics. Bi$_2$O$_2$Se nanowires synthesized by chemical vapor deposition method presented a high field-effect mobility up to ~1.34×10$^4$ cm$^2$V$^{-1}$s$^{-1}$, and exhibited ballistic transport in the low back-gate voltage ($V_g$) regime where conductance plateaus were observed. When further increasing the electron density by increasing $V_g$, we entered the phase coherent regime and weak antilocalization (WAL) was observed. The spin relaxation length extracted from the WAL was found to be gate tunable, ranging from ~100 nm to ~250 nm and reaching a stronger spin-obit coupling (SOC) than the two-dimensional counterpart (flakes).**

**We attribute the strong SOC and the gate tunability to the presence of a surface accumulation layer which induces a strong inversion asymmetry on the surface. Such scenario was supported by the observation of two Shubnikov-de Haas oscillation frequencies that correspond to two types of carriers, one on the surface, and the other in the bulk. The high-quality $Bi_2O_2Se$ nanowires with a high mobility and a strong SOC can act as a very prospective material in future spintronics.**

High-mobility nanowires with strong spin-orbit coupling (SOC) are compelling building blocks for semiconductor spintronics[1,2] and spin-based quantum information technology[3], as SOC enables an external electric field as an powerful knob to manipulate electron spins. Meanwhile, hybrid superconductor-semiconductor nanowire systems with strong SOC are attractive platforms for the implementation of spinless *p*-wave superconductivity[4,5], which serves as hosts for Majorana bound states to achieve topological quantum computing[6,7]. Outstanding transport properties of nanowires, such as a high-level tunability of the carrier density and a low-level disorder scattering, also play a fundamental role in the realization of these proposals.

$Bi_2O_2Se$, an emerged layered semiconductor formed by alternative layers of $[Bi_2O_2]^{2n+}$ and $[Se]^{2n-}$ layers, exhibits excellent electronic properties due to its moderate bandgap (~0.8 eV), small electron effective mass (~0.14 $m_e$), together with ultra-high mobility[8,9] (reaching 160,000 $cm^2V^{-1}s^{-1}$ in strain-free flakes[10]), making it a very attractive electronic[11] and optoelectronic[12,13-14] material. Furthermore, a strong SOC (spin relaxation length $L_{so}$ ~ 150-250 nm) has been found through magneto-transport measurements[15,16]. In view of these attractive properties, combined with its air stability and easy accessibility[17], much attention has been paid on two-dimensional (2D) $Bi_2O_2Se$ flakes. However, low-dimensional $Bi_2O_2Se$ systems, such as nanowires and quantum dots, have been less explored[18,19]. Here, high-quality $Bi_2O_2Se$ nanowires were synthesized and fundamental electrical transport characterizations were investigated to reveal its great transport properties and further application potentials in spintronics. Specifically, a high field-effect mobility up to ~$1.34\times10^4$ $cm^2V^{-1}s^{-1}$, and a strong SOC that exceeds its 2D counterpart, with a tunable spin relaxation length ranging from ~100 nm to ~250 nm, was uncovered.

High-quality Bi$_2$O$_2$Se nanowires were synthesized by chemical vapor deposition method, based on facial (Bi)-catalyzed vapor-liquid-solid mechanism[20]. Self-assembled nanowires were free-standing with out-of-plane morphology on mica substrate. The bottom-up synthesized nanowires have a typical width of 100~300 nm, an average thickness of ~10 nm, and a length up to 10 µm. Subsequent to the growth, the nanowires were transferred to SiO$_2$/Si substrate via a polymer assisted clean transfer method. A back-gate voltage $V_g$ could be applied through the degenerately doped Si substrate and the 300-nm-thick SiO$_2$. Electrode patterns were fabricated by electron-beam lithography, and Ti/Au (5/100 nm) contacts were deposited by electron-beam evaporation. A soft plasma cleaning was performed prior to the metal deposition to ensure good contact. Figure 1(a) shows a scanning electron microscope (SEM) image of the measured device with a channel length $L \sim 3.72$ µm, a width $W \sim 260$ nm, and a thickness of ~10 nm confirmed by atomic force microscopy measurement. Figure 1(b) shows a zoom-in of the nanowire. Transport measurements were performed in a $^3$He/$^4$He dilution refrigerator, with a magnetic field $B$ applied perpendicular to the substrate. The two-terminal conductance (resistance) was measured using a standard low-frequency lock-in technique.

Field-effect transistor (FET) measurements were first carried out in order to determine the electron mobility. Figure 1(c) shows the measured conductance of the device as a function of $V_g$ at a temperature of 2 K. As can be seen, the conduction channel remains closed at zero back-gate voltage and the pinch-off threshold voltage ($V_{th}$) is positive, which reveals that the fermi level of the nanowire resides in the semiconducting gap, displaying intrinsic semiconductor transistor characteristics.

The field-effect mobility can be extracted from the fitting of the transfer characteristics. The modified mobility expression incorporating the contribution of contact resistances is employed to precisely obtain the mobility[21],

$$G(V_g) = \left(R_s + \frac{L^2}{\mu C_g(V_g - V_{th})}\right)^{-1} \tag{1}$$

where µ is the mobility, $C_g$ is the capacitance, and $R_s$ is the series resistance. The mobility µ can be determined by fitting the measured conductance curve shown in Fig. 1(c). To do so, $C_g$ is precisely obtained using 2D Finite Element Method (FEM) simulations[22] with a rectangular cross-section of the nanowire. The obtained capacitance per unit length is $C_g^L$ ~66 pF/m, which is 2.5 times of that obtained from an analytical parallel plate model. From the good fitting as shown by the orange solid line in Fig. 1(c), we obtain $R_s$ ~ 809 Ω, $V_{th}$ ~ 6.9 V and a field-effect mobility, μ ~ $1.34 \times 10^4$ $cm^2V^{-1}s^{-1}$. The ultra-high mobility enables the nanowire to display distinct Shubnikov-de Haas (SdH) oscillations in magnetic fields, as shown later. Noteworthy, the high mobility in $Bi_2O_2Se$ stems from the unique electron conducting mechanism in the layered structures. The electron conduction channels predominantly locate in the $[Bi_2O_2]^{2n+}$ layers spatially away from $V_{Se}$ (Se-vacancy) defect states that locate in the $[Se]^{2n-}$ layers. Such a spatial separation can strongly suppress the scattering and give rise to the large electron mobility[23].

From the measured transfer characteristic near the pinch-off voltage, well-developed step-like features can be observed, revealing the population of confined 1D sub-bands and ballistic transport in the nanowire [Fig. 2(a)]. From Landauer formalism, the total conductance through the device with multiple conduction channels reads as $G = G_0 \sum_i T_i$, where $T_i$ is the transmission associated with the $i$-th channel. In the ideal situation without reflection, $T_i$ =1 and each sub-band contributes a quantum unit of conductance, $G_0 = 2e^2/h$ ($e$ is the elementary charge, and $h$ is the Planck's constant). As can be seen from Fig. 2(a), the conductance increases in steps of approximately $0.3G_0$ ($T_i$~0.3) rather than quantized values. We attribute the deviations from the quantized plateaus to the interface scattering unintentionally formed in the contact interface, which behaves as a potential barrier suppressing the conductance to be well below quantized values[24]. Note that the Fabry-Perot oscillations superimposed on the plateaus further suggest the ballistic transport through the nanowire and the existence of potential barriers. The Fabry-Perot oscillations become more evident as decreasing the temperature down to 10 mK, as shown in the inset of Fig. 2(a). As displayed in Fig. 2(b), at a finite dc-bias, diamond-like structures emerge due to the detuning between the source and drain chemical potentials, which provides an immediate measure of the energy separation between neighboring sub-bands. The ballistic transport can also be observed in another nanowire device with a longer channel length ($L$ ~6 μm and $W$ ~180 nm), as shown in the Supplementary material.

To gain more insight into the observed ballistic transport in the nanowires, generalized Landauer formalism was employed to better match the experimental results. Taking thermal and disorder broadening effect into account[25], the Landauer formula takes the following form,

$$G = \frac{2e^2}{h} \sum_i \int T_i(E) F(E) G(E) dE \qquad (2)$$

where $F(E)$ and $G(E)$ represent the thermal and disorder broadening function, respectively.

$$F(E) = \frac{1}{4k_B T} \operatorname{sech}^2 \left( \frac{E - E_F}{2k_B} \right) \qquad (3)$$

$$G(E) = \frac{1}{c\sqrt{2\pi}} \exp \left( \frac{-(E - E_F)^2}{2c^2} \right) \qquad (4)$$

where $k_B$ is Boltzmann's constant and $c$ is the variance of Gaussian distribution function. The relation between Fermi energy $E_F$ and $V_g$ can be established from the following identity,

$$N_{1D} = C_g^L (V_g - V_{th})/e = \frac{2\sqrt{2m^*}}{\pi \hbar} \sum_i \sqrt{E_F - E_i} \qquad (5)$$

where the right side denotes the total quantum states in 1D case, and the middle represents the number of electrons tuned by the gate. Note that $C_g^L$ is the capacitance in unit length, $m^*$ is the effective mass of $Bi_2O_2Se$ ($m^* \sim 0.14 m_e$). From the fitting to the experimental data as shown in Fig. 2(a), we get $C_g^L$ = 24 pF/m, $\Delta E_{i,i+1}$ ($i$ = 1, 2, 3) = 15, 12, 13 meV, $T_i$ ($i$ = 1, 2, 3) = 0.3, 0.35, 0.4, $c$ = 0.5 meV. The sub-band energy differences $\Delta E_{i,i+1}$ show consistence with that indicated by the diamond-like structures in Fig. 2(b), and the derived transmissions $T_i$ are consistent with the height of the conductance plateaus.

It should also be noted that the $C_g^L$ obtained here is smaller than half of that obtained from the FEM simulation shown above, which indicates that the carrier distribution near pinch-off voltage differs from that at higher $V_g$. As commonly found, a surface accumulation layer is present in nanowire systems such as InAs[26], so it is reasonable to assume the presence of an accumulation layer[19,27] in our Bi$_2$O$_2$Se nanowires which probably originates from the surface band bending induced by Se vacancies and thus accumulation of absorbates during the growth and especially in the polymer transfer process. Near pinch-off, the carriers seem to concentrate at the top accumulation layer, whereas at higher $V_g$ the carriers tend to populate both the accumulation layer and the inner bulk. We attribute the reduction of capacitance to the screening effect and the stronger back-gate tunability of carriers in bottom inner bulk rather than the top surface accumulation layer. In addition, one important remark should be made here on the ballistic transport in our micro-scale nanowires. In the low-$V_g$ regime, the lower electron concentration leads to larger electron wave length, which enables electrons to circumvent the defects (Se$_{Bi}$ in [Bi$_2$O$_2$]$^{2n+}$ layers), so that the transport can be easily ballistic. When increasing the electron concentration and decreasing the electron wave length by elevating the gate voltage, the intralayer scattering (Se$_{Bi}$ in [Bi$_2$O$_2$]$^{2n+}$ layers) and even the interlayer scattering (V$_{Se}$ in [Se]$^{2n-}$ layers) will come into play, and then we enter the phase coherent regime as we will discuss below.

When we elevate the back-gate voltage, owing to the decrease of the electron wave length, the scattering events emerge and quantum interference develops. The constructive interference of time reversed, closed electron trajectories can induce negative quantum correction to the classical Drude conductance, known as weak localization. If SOC is present, a destructive interference yields a positive conductance correction, known as weak antilocalization (WAL)[28-31].

Figure 3(a) displays the measured low-field magnetoconductance $\Delta G = G(B) - G(B=0)$ at $V_g = 14$ V at different temperatures. A magneto-conductance peak was clearly observed, indicating a WAL effect in the entire range of temperature (2-24 K) considered, and the peaks become broadened with increasing temperature. Note that the oscillations at low temperatures result from universal conductance fluctuations (UCFs). We can analyze the data based on the Hikami-Larkin-Nagaoka (HLN) quantum interference theory[32] with the conductance correction given by,

$$\Delta G(B) = -\frac{e^2}{\pi h}\left[\frac{1}{2}\Psi\left(\frac{B_\varphi}{B}+\frac{1}{2}\right)+\Psi\left(\frac{B_e}{B}+\frac{1}{2}\right)-\frac{3}{2}\Psi\left(\frac{(4/3)B_{SO}+B_\varphi}{B}+\frac{1}{2}\right)-\frac{1}{2}\ln\left(\frac{B_\varphi}{B}\right)\right.$$
$$\left.-\ln\left(\frac{B_e}{B}\right)+\frac{3}{2}\ln\left(\frac{(4/3)B_{SO}+B_\varphi}{B}\right)\right] \quad (6)$$

where $\Psi(x)$ is the digamma function, and $B_j$ ($j = \varphi$, so, e) are the characteristic magnetic fields and are related to the characteristic lengths via $B_j = \hbar/(4eL_j^2)$, with $L_\varphi$ being the phase coherence length, $L_{so}$ the spin relaxation length and $L_e$ the electron mean free path. To extract these relevant characteristic lengths, fitting is performed to the magnetoconductance curves, as shown by the solid lines in Fig. 3(a). The extracted characteristic lengths as a function of temperature are plotted in Fig. 3(b). It can be seen that $L_{so}$ and $L_e$ show a weak dependence on temperature other than small fluctuations, while $L_\varphi$ follows a power law dependence, $L_\varphi \sim T^{-0.41}$, which is consistent with the Nyquist dephasing mechanism by electron-electron scattering with small energy transfers[33]. From the fitting, the extracted $L_{so}$ is around 100 nm, which is smaller than the values reported from 2D $Bi_2O_2Se$ flakes[15], corresponding to a stronger SOC in the nanowires.

We further investigated the gate-voltage dependence of the WAL effect. Figure 3(c) displays the low-field magnetoconductance for different $V_g$ at $T=2$ K. Here, universal conductance fluctuations (UCFs) imposed on the WAL curves were generally observed, indicating phase coherent transport in the nanowires. Note that the conductance curves shown in Fig. 3(c) are the results of averaging several curves around a given $V_g$ in order to avert the influence of UCFs to some extent. We then fit the curves using the HLN formula to extract the gate dependence of the relevant characteristic lengths, as shown by the solid lines in Fig. 3(c). The results are plotted in Fig. 3(d). $L_\varphi$ exhibits a strong dependence on $V_g$. By varying $V_g$ from 40 V to 14 V, $L_\varphi$ decreases from 500 nm to 185 nm, suggesting an increasing dephasing for a lower electron density as a result of enhanced electron-electron interaction and reduced Coulomb screening[31]. It is worth noting that $L_{so}$ (i.e., the SOC strength) shows a gate tunability from ~250 to 100 nm when varying $V_g$ from 40 to 14 V.

We next present an elaborate analysis to illustrate the observed strong SOC and the gate tunability. In $Bi_2O_2Se$, the SOC can result from the intrinsic spatial inversion asymmetry between $[Bi_2O_2]^{2n+}$ layers and $[Se]^{2n-}$ layers, or from extrinsic potential naturally formed in the growth and device fabrication process[34]. The strong SOC indicates a strong inversion symmetry breaking in our nanowires. Under our assumption, the surface accumulation states provide a strong structure inversion asymmetry in the nanowires, causing an enhanced Rashba SOC. Whereas the inner bulk of the nanowire possesses negligible inversion asymmetry and thus contributes a relatively small part to the SOC. In addition, the single back gate employed here is mainly used to tune the electron density rather than the electric field piercing through the nanowire. Consequently, with increasing $V_g$, the contribution of surface accumulation carriers to the conduction electrons is weakened and the contribution from the inner bulk is weighted. Hence, the averaged SOC, which is stronger for surface accumulation states, is also weakened, consistent with our observations.

As mentioned previously, the high mobility enables the observation of SdH oscillations[35], by which the above scenario can be supported. Figrue 4(a) shows the magneto-resistance of the nanowire measured at $T = 2$ K and $V_g = 60$ V. Again, the cusp feature around zero magnetic field originates from the WAL effect, as shown in Fig. 4(b). The fitting to HLN theory gives characteristic lengths $L_j$ ($j = \varphi$, so, e) = 583, 274, 109 nm, respectively. In strong perpendicular magnetic fields, prominent SdH oscillations can be recognized. In Fig. 4(c), we show the variations of the resistance, $\Delta R$, vs. $1/B$ after subtracting a polynomial background. The corresponding fast Fourier transform (FFT) spectra of the SdH oscillations are presented in Fig. 4(d), and two oscillation frequencies $F_1 = 89.9$ T and $F_2 = 143.8$ T can be identified. Considering the Onsager relationship, $S_F = 2\pi^2 n = (4\pi^2 e/h)F$, where $F$ is the SdH oscillation frequency and $n$ the carrier density, the two SdH frequencies generate two carrier densities $n_1 = 4.35 \times 10^{12}$ cm$^{-2}$ and $n_2 = 6.95 \times 10^{12}$ cm$^{-2}$, and the total sheet carrier density $n = n_1 + n_2 = 11.3 \times 10^{12}$ cm$^{-2}$, which is comparable with the electron density obtained from FET results [$n = C_{gs}/e \times (V_g - V_{th})$ where $C_{gs}$ is the FEM capacitance per unit area, and $n \sim 8.43 \times 10^{12}$ cm$^{-2}$] . Note that the two frequencies observed here show similarity from that inuduced by SOC spin-splitted bands. From the WAL analysis as shown in Fig. 4(b), a spin relaxation length $L_{so} \sim 270$ nm corresponds to a Rashba parameter $\alpha \sim 2 \times 10^{-12}$ eV·m ($\alpha = \hbar^2/m^* L_{so}$, where $m^* = 0.14$ $m_e$), which tranlates to a spin-orbit energy of $\Delta \sim 4$ μeV ($\Delta = m^*\alpha^2/(2\hbar^2)$). However, if we consider zero-field spin-splitting induced by Rashba SOC[36], $\alpha = (\hbar^2/m^*)\sqrt{\pi/2} \cdot (\Delta n/\sqrt{n})$, where $\Delta n = n_2 - n_1$, we obtain $\alpha \sim 2.5 \times 10^{-11}$ eV·m which is one order of magnitude higher than that obtained from WAL results. Hence, we rule out the posibility of Rashba SOC induced zero-field spin-splitting and attribute the two frequencies to the carries at the surface accumulation layer and in the inner bulk.

In addtion, a multimode Fabry-Perot conductance oscillation was observed in a short channnel nanowire with a length of ~320 nm and a width of ~155 nm, as presented in the Supplementary material. It indicates that the high-mobility $Bi_2O_2Se$ nanowires can act as coherent electron waveguide resonator as found in carbon nanotubes[37] and InAs nanowires[38].

In conclusion, we synthesized $Bi_2O_2Se$ nanowires and performed systematic electrical transport investigations. A high mobility up to $\mu \sim 1.34 \times 10^4$ cm$^2$V$^{-1}$s$^{-1}$ and a gate-tunable SOC that exceeds the 2D flakes were achieved. The nanowire presents ballistic transport in the low back-gate voltage regime and enters the phase coherent transport when elevating the back-gate voltage where WAL was observed. The ultra-high mobility enables the observation of SdH oscillations, through which we identified two types of carries, one attributed to the surface accumulation layer and the other the inner bulk. The former contributes the strong inversion asymmetry and further derives the strong SOC in our nanowires. Therefore, our high-quality $Bi_2O_2Se$ nanowires present high mobility, strong SOC, and gate tunability, making it a prospective material in future electronics and spintronics, and a potential platform for realizing helical states and further for topological quantum computation.

SUPPLEMENTARY MATERIAL

See supplementary material for additional figures and text.

ACKNOWLEDGEMENTS

This work was supported by the National Basic Research Program of China from the MOST Grants No. 2017YFA0304700 and 2016YFA0300601, by the NSF China Grants No. 12074417, 92065203, 11874406, 11774405, 92164205 and 11527806, by the Strategic Priority Research Program B of Chinese Academy of Sciences, Grants No. XDB28000000 and XDB33010300, by the Synergetic Extreme Condition User Facility sponsored by the National Development and Reform Commission, and by the Innovation Program for Quantum Science and Technology (Grant No. 2021ZD0302600).

AUTHOR DECLARATIONS

Conflict of Interest

      The authors have no conflicts to disclose.

Author Contributions

K.Z., H.L. and C.T. contributed equally to this work.

**Kui Zhao**: Conceptualization (equal); Data curation (equal); Formal analysis (equal); Investigation (equal); Writing-original draft (lead). **Huaiyuan Liu**: Conceptualization (equal); Formal analysis (equal); Investigation (equal). **Congwei Tan**: Conceptualization (equal); Data curation (equal); Investigation (equal). **Jianfei Xiao**: Data curation(supporting); Investigation (equal). **Jie Shen**: Data curation (supporting). **Guangtong Liu**: Data curation (supporting). **Hailin Peng**: Investigation (supporting); Funding acquisition (equal); Supervision (equal). **Li Lu**: Investigation (supporting); Funding acquisition (equal); Supervision (equal). **Fanming Qu**: Conceptualization (equal); Funding acquisition (equal); Investigation (equal); Supervision (equal); Writing – review & editing (lead).

DATA AVAILABILITY

The data that support the findings of this study are available within the article and its supplementary material and from the corresponding author upon reasonable request.

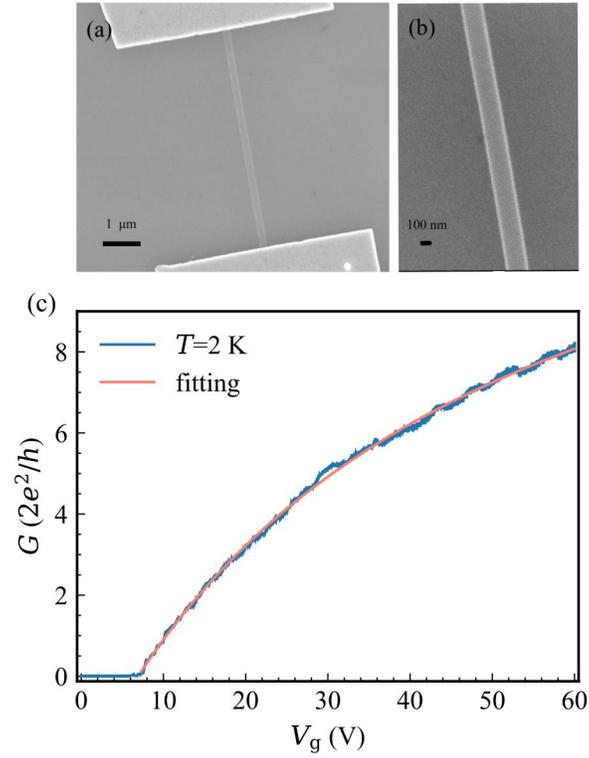

FIG. 1. (a) SEM image of the measured device, a $Bi_2O_2Se$ nanowire contacted to two Ti/Au electrodes. (b) Zoom-in of the nanowire in (a). (c) Conductance of the nanowire as a function of the back-gate voltage at $T = 2$ K (blue curve) measured at a source-drain bias of 10 mV. Field-effect mobility is extracted from a fitting (orange curve) using Eq. (1).

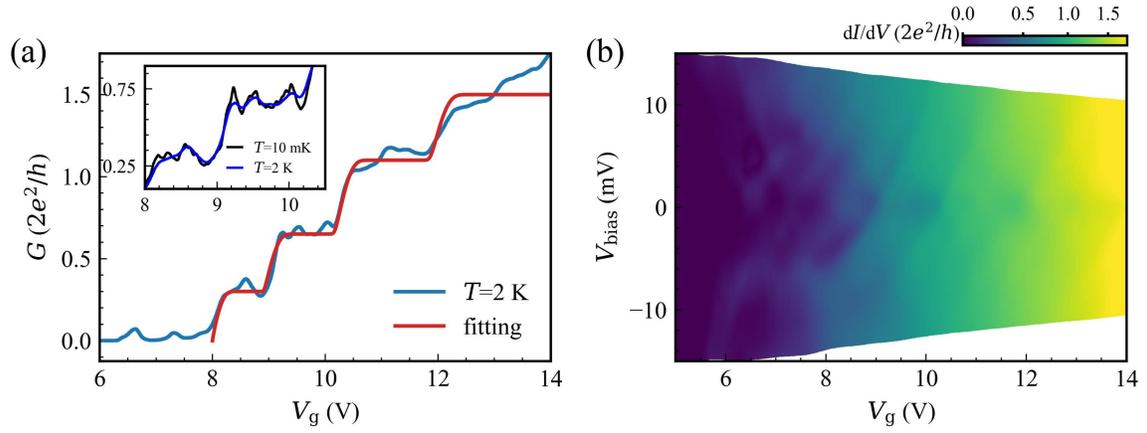

FIG. 2. (a) Zoom-in of the low-$V_g$ regime in Fig. 1(a). The two-terminal conductance (channel length 3.72 µm) was measured using a quasi-four terminal configuration at zero magnetic field, where a 17 Hz AC excitation voltage of 100 µV was supplied, and the current ($I$) and the voltage drop ($V$) on the device were recorded. Conductance plateaus can be clearly recognized (blue line). The red curve is a fitting to the data using Eq. (2). The inset shows the zoom-in of Fabry-Perot oscillations imposed on the conductance steps. (b) Differential conductance d$I$/d$V$ as a function of $V_g$ and dc bias voltage $V_{bias}$, showing diamond-shaped structures.

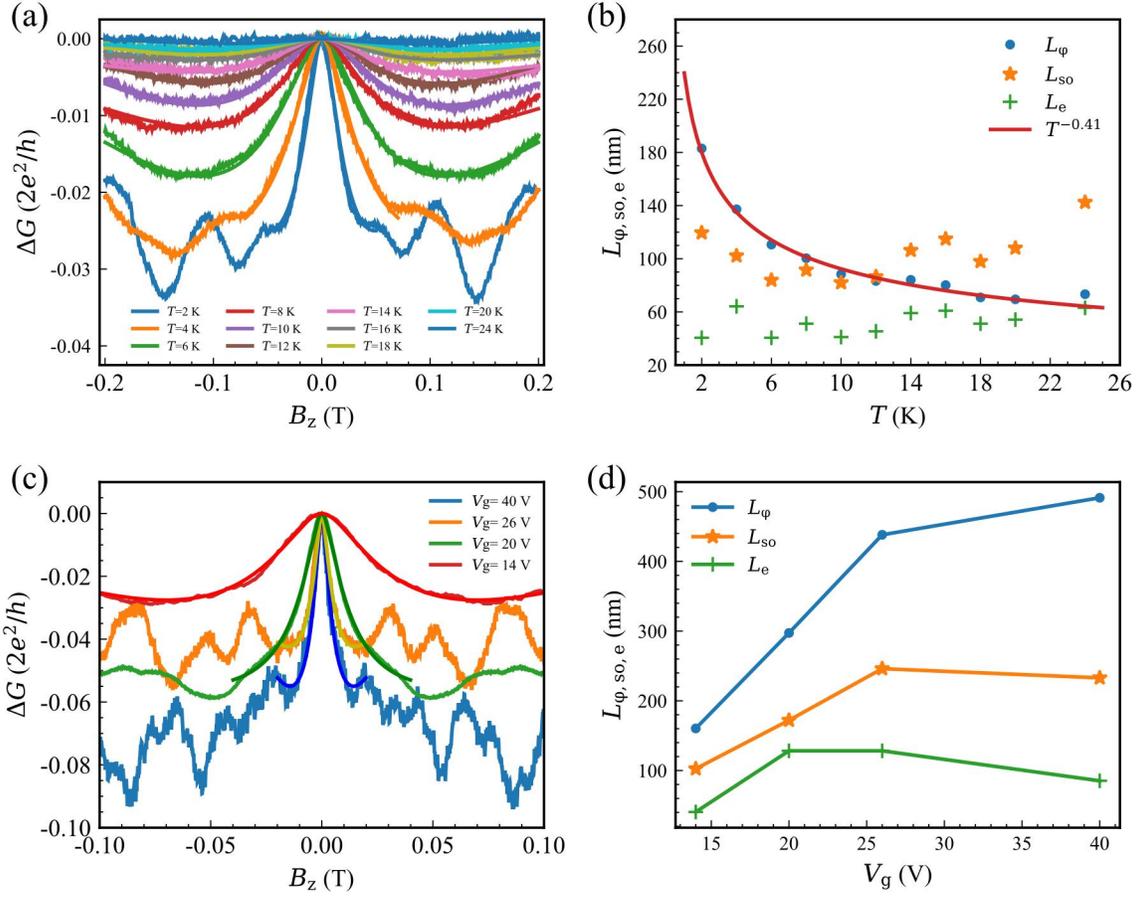

FIG. 3. (a) Low-field magneto-conductance $\Delta G$ measured at $V_g = 14$ $V$ at different temperatures. The solid lines are the fittings of the measured data using the HLN theory. (b) Characteristic lengths $L_j$ ($j = \varphi, so, e$) as a function of temperature, extracted from the fitting in (a). The red curve shows a power law fitting of $L_\varphi$. (c) Magneto-conductance measured at $T = 2$ K and at various $V_g$. Solid lines are fittings using the HLN theory. (d) Characteristic lengths $L_j$ ($j = \varphi, so, e$) as a function of $V_g$, extracted from the fitting in (c).

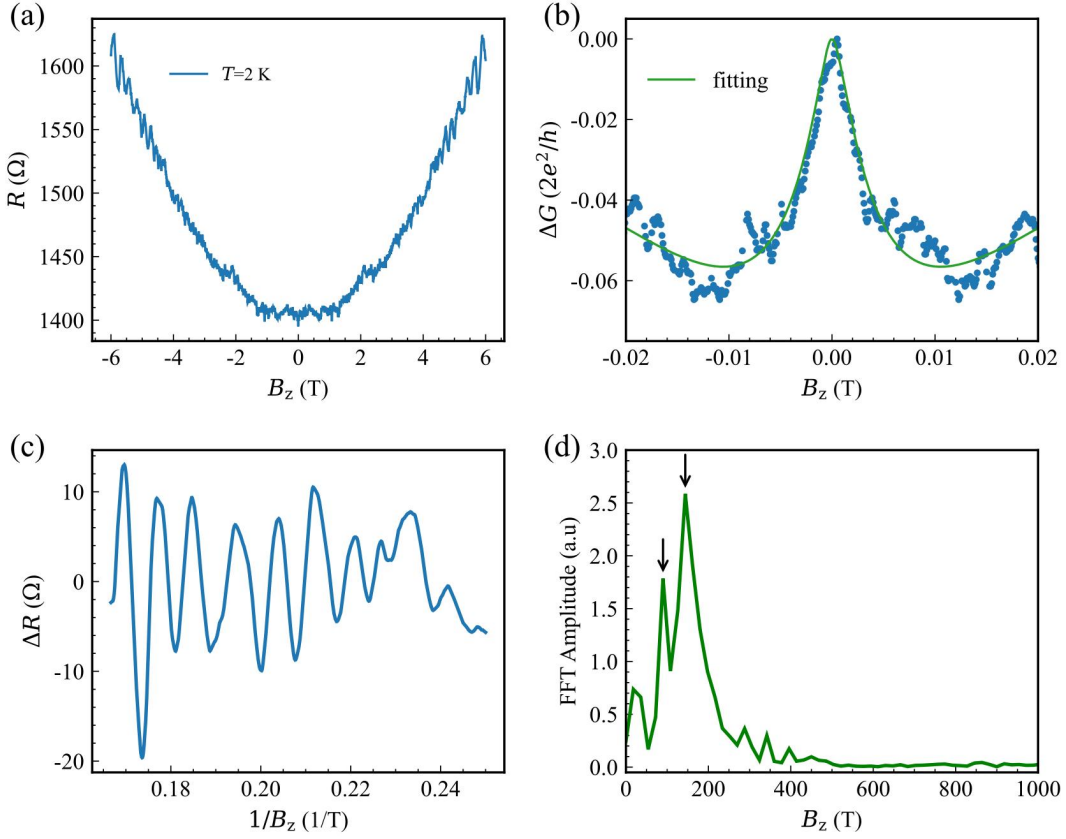

FIG. 4. Magneto-resistance (MR) of the nanowire measured at $T = 2$ K and $V_g = 60$ V. (a) MR shows WAL in low magnetic field and SdH oscillations in high magnetic field. (b) Low-field magneto-conductance $\Delta G$ in (a), showing WAL effect. The solid curve is the fitting using the HLN theory, with extracted characteristic lengths $L_j$ ($j = \varphi, so, e$) = 583, 274, 109 nm, respectively. (c) SdH oscillations obtained by subtracting a polynomial background from the MR in (a). (d) FFT of the data in (c), displaying two distinct frequencies as indicated by the black arrows.

# Supplementary Material for
# Bi$_2$O$_2$Se nanowires presenting high mobility and strong spin-orbit coupling

Kui Zhao[1,2], Huaiyuan Liu[1,2], Congwei Tan[3], Jianfei Xiao[1,2], Jie Shen[1,4], Guangtong Liu[1,4], Hailin Peng[3,a)], Li Lu[1,2,4,a)], and Fanming Qu[1,2,4,a)]

[1]*Beijing National Laboratory for Condensed Matter Physics, Institute of Physics, Chinese Academy of Sciences, Beijing 100190, China*

[2]*School of Physical Sciences, University of Chinese Academy of Sciences, Beijing 100049, China*

[3]*Center for Nanochemistry, Beijing Science and Engineering Center for Nanocarbons, Beijing National Laboratory for Molecular Sciences, College of Chemistry and Molecular Engineering, Peking University, Beijing 100871, China*

[4]*Songshan Lake Materials Laboratory, Dongguan, Guangdong 523808, China*

a) Authors to whom correspondence should be addressed: hlpeng@pku.edu.cn; lilu@iphy.ac.cn; fanmingqu@iphy.ac.cn

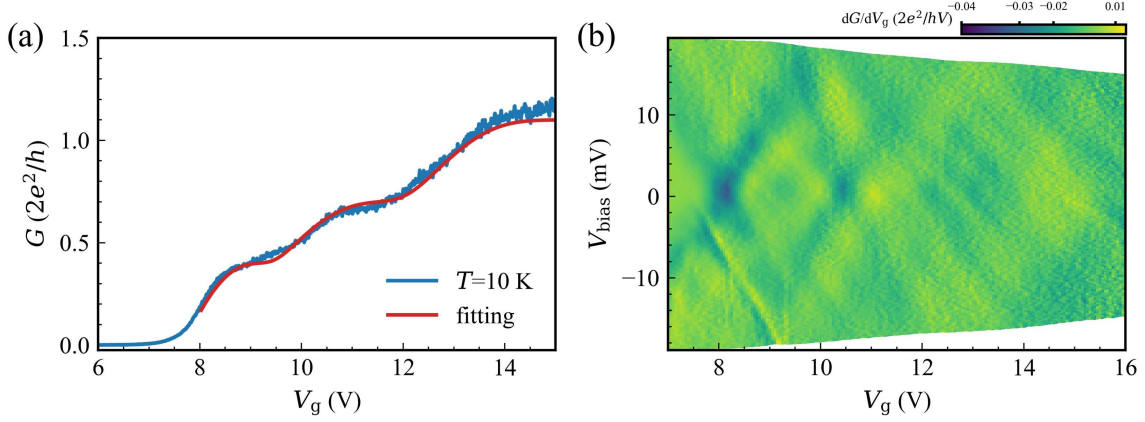

FIG. S1. Conductance plateaus observed in a device with longer channel length device ($L$ ~6 μm and $W$ ~180 nm). (a) Conductance plateaus are clearly observed (blue line). The red curve is a fitting to the data using Eq. (2) in the main text, with fitting parameters $C_g^L$ = 13 pF/m, $\Delta E_{i,\,i+1}(i = 1, 2)$ =19, 17 meV, $T_i$ ($i$ = 1, 2) = 0.4, 0.3, $c$ = 3 meV. (b) Transconductance $dG/dV_g$ as a function of $V_g$ and dc bias voltage $V_{bias}$, showing diamond-shaped conductance plateaus.

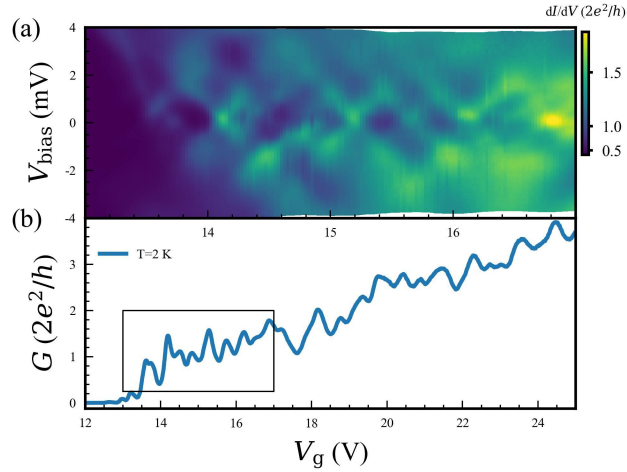

FIG. S2. Fabry-Perot oscillations observed in another short channel device ($L$ ~320 nm and $W$ ~ 155 nm). (a) Characteristic chessboard pattern of the differential conductance as a function of $V_g$ and $V_{bias}$, resulting from Fabry-Perot resonances at $T$ = 2 K. (b) Differential conductance oscillations as a function of $V_g$. The boxed region corresponds to the gate voltage regime in (a). For this device with a short channel length, strong Fabry-Perot resonances are imposed on the conductance plateaus with a height of ~$2e^2/h$.

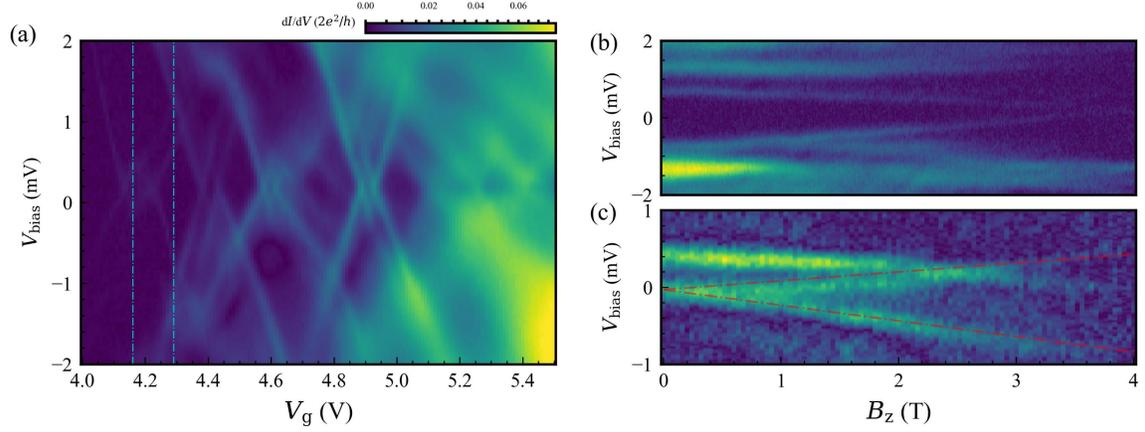

FIG. S3. Coulomb blockade phenomena observed in a $Bi_2O_2Se$ quantum dot device. (a) Stability diagram as a function of $V_{bias}$ and $V_g$ at $T = 2$ K. (b) Magnetic-Field evolution of differential conductance at $V_g = 4.16$ V along the left cyan line cut in (a). (c) Magnetic-Field evolution of differential conductance at $V_g = 4.26$ V along the right cyan line cut in (a). By fitting the peak shift in the full magnetic field range considering the spin-1/2 Zeeman energy term $E_z = 1/2|g|\mu_B B$, we can extract the g-factor associated with these two quantum levels as $|g| \sim 5$-$7$. Note that the g-factor could be level-dependent or geometry-dependent.